\documentstyle[12pt,aps,prd,preprint]{revtex}
\begin{document}
\title{Mode-Coupling in Rotating Gravitational Collapse of a Scalar Field}
\author{Shahar Hod}
\address{The Racah Institute for Physics, The
Hebrew University, Jerusalem 91904, Israel}
\date{\today}
\maketitle

\begin{abstract}

We present an {\it analytic} study of the mode-coupling phenomena for
a scalar field propagating on a {\it rotating} Kerr background.
Physically, this phenomena is caused by the dragging of reference frames, due to
the black-hole (or star's) rotation. We find that 
different modes become {\it mixed} during the evolution and the
asymptotic late-time tails are dominated by a mode which, in general,
has an angular distribution different from the original one. We show
that a rotating Kerr black hole becomes `bald' {\it slower} than a
spherically-symmetric Schwarzschild black hole.
\end{abstract}

\section{Introduction}\label{Sec1}

The radiative tail of gravitational collapse decays with time leaving
behind a Kerr-Newman black hole characterized solely 
by the black-hole mass, charge and angular-momentum. This is the
essence of the {\it no-hair conjecture}, introduced by Wheeler in the
early 1970s \cite{Whee}. The relaxation process of neutral fields 
was first analyzed by Price \cite{Price} for a nearly spherical collapse. Price
demonstrated that the fields decay asymptotically as an inverse power
of time. Physically, these inverse power-law tails are associated with
the backscattering of waves off the effective curvature
potential at asymptotically far regions \cite{Thorne,Price}.

The existence of these inverse power-law tails, which characterize the 
asymptotic late-time
evolution of test fields on a Schwarzschild background was later
verified by other authors \cite{Leaver,Gundlach1,Andersson,Barack1}
using independent analyses. The late-time evolution of natural test fields on
different spherically symmetric spacetimes was studied 
in Refs. \cite{Bicak,Gundlach1,Ching}. In addition, the late-time tails of
gravitational collapse were demonstrated by full numerical
computations \cite{Gundlach2,BurOr}.
The physical mechanism by which 
a charged black hole, which is formed during a gravitational collapse 
of a charged matter, dynamically sheds its {\it charged} hair was first studied
analytically in \cite{HodPir1,HodPir2}, and a full non-linear numerical analysis
was performed in \cite{HodPir3}.

However, all the above mentioned analyses were restricted to spherically
symmetric backgrounds (in particular, to the 
Schwarschild and Reissner-Nordstr\"om black holes). 
On the other hand, astrophysical objects are usually rotating. 
Hence, the physical process of stellar core 
collapse to form a black hole is expected
to be {\it non-}spherical in nature. 
The corresponding problem of wave dynamics outside a 
realistic, {\it rotating} black hole is much more complicated due to
the lack of spherical symmetry. Originally, this problem was addressed
numerically by Krivan et. al. \cite{Krivan1,Krivan2}, followed by Ori \cite{Ori}
and Barack \cite{Barack2} who provided preliminary analytical results
for the evolution of a test scalar field on a Kerr background.

Evidently, the most interesting situation from a physical point of
view is the propagation of {\it gravitational} waves 
on a {\it rotating} Kerr background. The evolution of higher-spin fields (and in
particular gravitational perturbations) outside a realistic, 
rotating black hole was studied analytically only recently
\cite{Hod}. This was done by analyzing the asymptotic late-time
solutions of Teukolsky's master equation \cite{Teukolsky1,Teukolsky2},
which governs the evolution of gravitational, electromagnetic,
neutrino and scalar perturbations fields on Kerr spacetimes. 

The main result presented in Ref. \cite{Hod} is the existence 
of inverse {\it power-law} tails at the asymptotic regions of
timelike infinity $i_+$, null infinity $scri_+$ and at the
black-hole outer horizon $H_+$ (where the
power-law behaviour is multiplied by an oscillatory term, caused by 
the dragging of reference frames at the event horizon).
This late-time behaviour is similar to the 
late-time behaviour of massless fields on a 
{\it spherically}-symmetric Schwarzschild background. 

However, there is one important feature (besides the oscillatory
behaviour along the black-hole horizon) which is unique to {\it rotating}
collapse: Active {\it coupling} of different modes. 
The phenomena of mode-coupling has been observed in 
numerical solutions of Teukolsky's equation \cite{Krivan1,Krivan2} 
(a similar phenomena is known in the case of rotating stars \cite{Kojima}). 
Physically, this phenomena is caused by the dragging of reference frames, due to
the rotation of the black hole (or the star). The issue of
mode-coupling was discussed in Ref. \cite{Hod}, where preliminary
considerations and results were presented. In this paper we 
give a {\it complete} mathematical analysis of this important
phenomena for the case of a scalar field propagating on a 
realistic Kerr background, and we focus on its physical implications. 
We consider {\it all} the various possible cases of 
mode-coupling, which result in different types of 
asymptotic late-time evolutions (different damping exponents).

We find it most convenient to use the basis of the 
(frequency-dependent) spheroidal wave
functions and carry out the analysis in the {\it frequency} domain; 
these functions allow us to separate the
$\theta$-dependence from the radial dependence, and, more important,
allow us to separate the different modes (different values of $l$) so 
each mode evolves {\it independently}. On the
other hand, our main objective is to study the {\it temporal} dependence of
the field's late-time behaviour; Therefore, 
at the end of the analysis we should integrate over all frequencies in the
frequency plane to obtain the explicit temporal 
dependence of the field. Thus, the {\it final} 
expression for the field is most naturally expressed in terms of 
the $\omega$-{\it independent} spherical harmonics.

The plan of the paper is as follows. In Sec. \ref{Sec2} we give a
short description of the physical system and summarize the main 
analytical results presented in Ref. \cite{Hod}. In Sec. \ref{Sec3} we discuss
the effects of rotation and the mathematical tools needed for
the physical analysis are derived. 
In Sec. \ref{Sec4} we study the active coupling of different
modes during a rotating gravitational collapse, with pure initial data.
In Sec. \ref{Sec5} we consider the coupling of different
modes, with generic initial data.
In the appendix we study the coupling 
of the real and the imaginary parts of the radial field.
We conclude in Sec. \ref{Sec6} with a summary of our analytical results.

\section{Review of Recent Analytical Results}\label{Sec2}

We consider the evolution of a massless scalar 
field outside a {\it rotating} star or a black hole.
The external gravitational field of a rotating object of
mass $M$ and angular-momentum per unit-mass $a$ is given by the
Kerr metric, which in Boyer-Lindquist coordinates takes the form

\begin{eqnarray}\label{Eq1}
ds^{2}&=& \left( 1 -{{2Mr} \over \Sigma}\right) dt^{2} +
\left( {{4Mar\sin^{2} \theta} \over \Sigma} \right) dtd \varphi-
{\Sigma \over \Delta} dr^{2} -\Sigma d\theta^{2} - \nonumber \\
&& \sin^{2} \theta \left( r^{2} +a^{2} +{{2Ma^{2}r\sin^{2} \theta} \over
    \Sigma} \right) d\varphi^{2} \  ,
\end{eqnarray}
where $\Sigma=r^{2}+a^{2}\cos^{2} \theta$ and $\Delta=r^{2} -2Mr
+a^{2}$ (we use gravitational units in which $G=c=1$).

In Boyer-Lindquist coordinates the wave equation for a massless scalar
field reads \cite{Brill}

\begin{eqnarray}\label{Eq2}
&& \left[{{(r^{2}+a^{2})^{2}} \over {\Delta}} 
-a^{2}\sin^{2}\theta \right]
{{\partial ^2 \psi} \over {\partial t^2}} 
+{{4Mar} \over {\Delta}} 
{{\partial ^2 \psi} \over {\partial t \partial \varphi}} 
+ \left[{{a^{2}} \over \Delta} -{1 
\over {\sin^{2} \theta}} \right] {{\partial ^2 \psi} 
\over {\partial \varphi ^2}} \nonumber \\
&& - {\partial \over {\partial r}} 
\left( \Delta {{\partial \psi} 
\over {\partial r}} \right) -{1 \over {\sin\theta}} 
{{\partial} \over {\partial
    \theta}} \left( \sin \theta{{\partial \psi} \over {\partial
    \theta}} \right)  =0\ \ .
\end{eqnarray}
Resolving the field in the form

\begin{equation}\label{Eq3}
\psi= (r^{2}+a^{2})^{-1/2} 
\sum\limits_{m= -\infty}^{\infty} 
\Psi^{m} e^{im \varphi}\ \ ,
\end{equation}
where $m$ is the azimuthal number, 
one obtains a wave-equation for each value of $m$:

\begin{equation}\label{Eq4}
B_{1}(r,\theta) {{\partial ^2 \Psi} \over {\partial t^2}} +
B_{2}(r,\theta) {{\partial \Psi} \over
{\partial t}} -
{{\partial ^2 \Psi} \over {\partial y^2}}+ B_{3}(r,\theta) \Psi - 
{{\Delta} \over {(r^{2}+a^{2})^2}} {1 \over {\sin\theta}} 
{{\partial} \over {\partial
    \theta}} \left( \sin \theta{{\partial \Psi} \over {\partial
    \theta}} \right)=0\ \ ,
\end{equation}
where the tortoise radial coordinate $y$ is defined 
by $dy={{r^{2}+a^{2}} \over \Delta} dr$.
The coefficients $B_{i}(r,\theta)$ are given by

\begin{equation}\label{Eq5}
B_{1}(r,\theta)=1-{{\Delta a^{2}\sin^{2}\theta} \over {(r^{2}+a^{2})^{2}}}\ \ ,
\end{equation}

\begin{equation}\label{Eq6}
B_{2}(r)= {{4iMmar} \over {(r^2+a^2)^2}}\ \ ,
\end{equation}
and

\begin{eqnarray}\label{Eq7}
B_{3}(r,\theta)&=&\Bigg [ 2(r-M) r(r^2+a^2)^{-1}  -m^{2} \left( {{a^{2}} 
\over \Delta} -{1 \over {\sin^{2}\theta}} \right) \Bigg] 
{\Delta \over {(r^2+a^2)^2}}\ \ .
\end{eqnarray}
The time-evolution of a wave-field 
described by Eq. (\ref{Eq4}) is given by

\begin{eqnarray}\label{Eq8}
\Psi (z,t) &=& 2\pi \int \int_{0}^{\pi} \Bigg\{ B_{1}(z') 
\Big [ G(z,z';t) \Psi _t(z',0)+G_t(z,z';t) 
\Psi (z',0) \Big] +  \nonumber \\ 
&& B_{2}(z') G(z,z';t) \Psi (z',0) \Bigg\} \sin\theta' d\theta'dy'\ \ ,
\end{eqnarray}
for $t>0$, where $z$ stands for $(y,\theta)$, and $G(z,z';t)$ is 
the (retarded) Green's function, which is defined by

\begin{eqnarray}\label{Eq9}
&&\left [ B_{1}(r,\theta) {{\partial ^2} \over {\partial t^2}} +
B_{2}(r,\theta) {{\partial} \over
{\partial t}} -
{{\partial ^2} \over {\partial y^2}}+ B_{3}(r,\theta) - 
{{\Delta} \over {(r^{2}+a^{2})^2}} {1 \over {\sin\theta}} 
{{\partial} \over {\partial
    \theta}} \left( \sin \theta{{\partial} \over {\partial
    \theta}} \right) \right ] G(z,z';t)=  \nonumber \\
&&  \delta (t) \delta(y-y') 
{{\delta(\theta - \theta')} \over {2\pi \sin\theta}} \ \ .
\end{eqnarray}
The Green's function (which is associated with the existence of a
branch cut in the complex frequency plane \cite{Leaver}, usually
placed along the negative imaginary $\omega-$ axis) 
was calculated in Ref. \cite{Hod} [see Eq. (31)
of \cite{Hod}]:

\begin{eqnarray}\label{Eq10}
G^C(z,z';t)& =&  \sum\limits_{l=|m|}^{\infty} 
{{ (-1)^{l+1} 2^{2l}} \over {\pi A^{2} [(2l+1)!]^2}} 
\int_{0}^{-i \infty} {{\Gamma(l+1-2i\omega M)} 
\over {\Gamma(-l-2i\omega M)}} \nonumber \\
&& \tilde \Psi_1(y,\omega) 
\tilde \Psi_1(y',\omega){S_{l}}(\theta,a\omega){S_{l}}(\theta',a\omega)
\omega ^{2l+1} e^{-i \omega t} d \omega\  ,
\end{eqnarray}
where the function $\tilde \Psi_1(y',\omega)$ is
given by (see Ref. \cite{Hod} for additional details):

\begin{equation}\label{Eq11}
\tilde \Psi_1 =Ar^{l +1} e^{i \omega r} M(l+1-2i\omega M ,2l +2, 
-2i \omega r)\  ,
\end{equation}
and $A$ is a normalization constant. This expression for the Green's
function is obtained using the large-$r$ (or equivalently, the law-$\omega$
approximation \cite{Hod}).
The functions $S_{l}^{m}(\theta,a\omega)$ are the spheroidal wave functions
which are solutions to the angular equation \cite{Flammer,Teukolsky2}

\begin{equation}\label{Eq12}
\left [  {1 \over {sin\theta}} 
{{\partial} \over {\partial
    \theta}} \left( \sin \theta{{\partial} \over {\partial
    \theta}} \right) +a^2\omega ^2 \cos^{2}\theta
 -{{m^{2}} \over
  {\sin^{2}\theta}} +{A_{l}^{m}} 
\right ]
{S_{l}^{m}}  =0  \  .
\end{equation}
For the $a\omega=0$ case, the eigenfunctions $S_{l}^{m}(\theta,a\omega)$
reduce to the well-known spherical harmonics
$Y_{l}^{m}(\theta,\phi)={S_{l}^{m}(\theta)e^{im\varphi}}$, and the
separation constants $A_{l}^{m}(a\omega)$ are 
simply $A_{l}^{m}=l(l+1)$.

\section{Rotation Effects}\label{Sec3}

The {\it rotational} dragging of reference frames, due to the rotation
of  the black hole (or the star) produces an active {\it coupling}
(during the gravitational collapse) between {\it different} modes.
Thus, even if the initial data is made of a pure mode (i.e., it
is characterized by a particular spherical 
harmonic $Y_{l}^{m}$) other modes would be 
generated dynamically during the evolution. 
Mathematically, the coupling of different modes is attributed to
the $\theta$-dependence of the spheroidal wave 
functions $S_{l}^{m}(\theta,a\omega)$ and of the
coefficient $B_{1}(r,\theta)$.

It is well known that the late-time behaviour of massless perturbations 
fields is determined by the backscattering from asymptotically {\it far}
regions \cite{Thorne,Price}. Thus, the late-time behaviour is dominated by the
{\it low}-frequencies contribution to the Green's function, for only low
frequencies will be backscattered by the small effective curvature 
potential (for $r \gg M$).
Thus, as long as the observer is situated far away from the 
black hole and the initial data has a considerable support only far away from 
the black hole, a {\it small}-$\omega$ approximation is sufficient in order
to study the asymptotic {\it late-time} behaviour of the field \cite{Andersson}.

If the angular equation (\ref{Eq12}) is written in the form

\begin{equation}\label{Eq13}
(L^{0}+L^{1})S_{l}^{m}=-A_{l}^{m}S_{l}^{m}\  ,
\end{equation}
with

\begin{equation}\label{Eq14}
L^{0}(\theta)={1 \over {\sin\theta}} 
{{d} \over {d
    \theta}} \left( \sin \theta{{d} \over {d
    \theta}} \right)  -{{m^{2}} \over {\sin^{2}\theta}}\  ,
\end{equation}
and

\begin{equation}\label{Eq15}
L^{1}(\theta,a\omega)=(a\omega)^2 \cos^{2}\theta\  ,
\end{equation}
it is immediately amenable to a perturbation treatment for small
$a\omega$. The spherical functions $Y_{l}^{m}$ are used as a
representation. They satisfy $L^{0}Y_{l}=-A_{l}^{(0)}Y_{l}$ 
with $A_{l}^{(0)}=l(l+1)$ (we suppress the index $m$ on $A_l$ and $Y_l$). 
For small $a\omega$ standard perturbation theory yields 
(see, for example, \cite{Schiff})

\begin{equation}\label{Eq16}
A_l=A_l^{(0)}+L^{1}_{ll}+\sum\limits_{k \neq l} {{|L^{1}_{lk}|}^2 \over
  {A_l^{(0)}-A_k^{(0)}}} + \cdots
\end{equation}
and

\begin{eqnarray}\label{Eq17}
S_l &=&Y_l+\sum\limits_{k \neq l} {L^{1}_{kl} \over
  {A_l^{(0)}-A_k^{(0)}}} Y_k + \nonumber \\
&& \bigg[\sum\limits_{k \neq l} \sum\limits_{n \neq l}
{L^{1}_{kn}L^{1}_{nl} \over {(A_l^{(0)}-A_k^{(0)})(A_l^{(0)}-A_n^{(0)})}} Y_k - 
\sum\limits_{k \neq l} {{L^{1}_{ll}L^{1}_{kl}} \over
  {(A_l^{(0)}-A_k^{(0)})^2}} Y_k \Bigg] +\cdots
\end{eqnarray}
where 

\begin{equation}\label{Eq18}
L^1_{lk} \equiv  \langle 0lm|L^1|0km \rangle \equiv
\int {Y_l^{m*}} L^1  {Y_k^m} d \Omega\  .
\end{equation}
Equation (\ref{Eq17}) implies that the black-hole rotation mixes
various spherical harmonics functions.

The spherical harmonics are related to the rotation
matrix elements of quantum mechanics \cite{CamMor}. Hence, standard
formulae are available for integrating the product of three such
functions. The one we need in order to evaluate the integral
Eq. (\ref{Eq18}) is 

\begin{equation}\label{Eq19}
\langle 0lm|\cos^2\theta|0km \rangle ={1 \over 3} \delta_{kl}+{2 \over 3}
\left({{2k+1} \over {2l+1}}\right)^{1/2}
\langle k2m0|lm\rangle \langle k200|l0 \rangle\  ,
\end{equation}
where $\langle j_1j_2m_1m_2|JM \rangle$ is a Clebsch-Gordan
coefficient. The well-known properties 
of the Clebsch-Gordan coefficients \cite{Abram}:

\begin{enumerate}
\item $\langle j_1j_2m_1m_2|JM \rangle$ vanishes unless $j_1, j_2$ and $J$ satisfy the
triangle selection rule $|j_1-j_2| \leq J \leq j_1+j_2$.

\item $\langle j_1j_200|J0 \rangle=0$ if $j_1+j_2+J$ is an odd 
integer (this is a parity conservation law).
\end{enumerate}
imply that $L^1_{lk} \neq 0$ for $l =k, k \pm 2$ (all other matrix
elements vanish). In general, if we choose the spherical harmonics as
a representation, so that

\begin{equation}\label{Eq20}
S_{l}(\theta,a\omega)= \sum\limits_{k=|m|}^{\infty} C_{lk}(a\omega)^{|l-k|} Y_{k}(\theta)\  ,
\end{equation}
then, to leading order in $(a\omega)^2$, the 
coefficients $C_{lk}(a\omega)$ take the form (see, for example, \cite{Schiff})

\begin{equation}\label{Eq21}
C_{lk}= \prod\limits_{n=l,l+2,\ldots}^{k-2} {{\langle 0n+2m|\cos^2\theta|0nm \rangle}
 \over
  {A_l^{(0)}-A_{n+2}^{(0)}}}\  ,
\end{equation}
if $k \geq l+2$ and $k-l$ is even, and

\begin{equation}\label{Eq22}
C_{lk}= \prod\limits_{n=k,k+2,\ldots}^{l-2} {{\langle 0nm|\cos^2\theta|0n+2m \rangle}
 \over
  {A_l^{(0)}-A_{n}^{(0)}}}\  ,
\end{equation}
if $k \leq l-2$ and $l-k$ is even. In addition, $C_{lk}=0$ if $|l-k|$ is
odd, and $C_{ll}=1$.

The time-evolution of a wave-field 
described by Eq. (\ref{Eq4}) is given by Eq. (\ref{Eq8}).
The coefficient $B_{1}(r,\theta)$ appearing in 
Eqs. (\ref{Eq4}) and (\ref{Eq8}) depend explicitly on the 
angular variable $\theta$ through the {\it rotation} of the black hole
(no such dependence exist in the non-rotating $a=0$ case).
To elucidate the coupling between different modes we should
evaluate the integral [see Eqs. (\ref{Eq5}) and (\ref{Eq8})]

\begin{equation}\label{Eq23}
\langle 0lm|\sin^2\theta|0km \rangle ={2 \over 3} \delta_{kl}-{2 \over 3}
\left({{2k+1} \over {2l+1}}\right)^{1/2}
\langle k2m0|lm\rangle \langle k200|l0 \rangle\  .
\end{equation}
Clearly, this integral vanishes unless $l=k,k \pm 2$. This, togather
with the expressions for the coefficients $C_{lk}$ imply that if a
mode $Y_{l^*}^m$ presents in the initial data, then all modes $Y_{l}^m$
with $|l^*-l|$ {\it even} (and $l \geq |m|$) will be present at late times.

As long as we are interested in the asymptotic late-time behaviour
(the $\omega \to 0$ limit) of the solution, we may use the approximation 

\begin{equation}\label{Eq24}
{\Gamma(l+1-2i\omega M) \over \Gamma(-l-2i\omega M)} \simeq 
2i\omega M(-1)^{l+1}l!^2\  ,
\end{equation}
which is valid for $\omega \to 0$. Thus, we obtain [see Eq. (\ref{Eq10})]

\begin{eqnarray}\label{Eq25}
G^C(z,z';t)& =&  \sum\limits_{l=|m|}^{\infty} 
{{2i M  } \over {\pi A^{2} [(2l+1)!!]^2}}  \nonumber \\
&& \int_{0}^{-i \infty} \tilde \Psi_1(y,\omega) 
\tilde \Psi_1(y',\omega) {S_{l}(\theta,a\omega)} {S_{l}(\theta',a\omega)}
\omega ^{2l+2} e^{-i \omega t} d \omega\  .
\end{eqnarray}

A note is needed here on the separation of the $r$ and $\theta$ variables in the Kerr
background. These variables are separated by
Teukolsky's equation only in the {\it frequency} domain. 
This is caused by the fact that
the spheroidal harmonics which are used for the separation 
depend {\it explicitly} on the temporal frequency $\omega$. On the
other hand, our main objective is to obtain expressions for the late-time behaviour of the
field in terms of the temporal variables ($t, u$, and $v$); 
at the end of the analysis we should recover the explicit {\it temporal}
dependence of the field by integrating over all frequencies in the
frequency plane (this would lead to the final answer which is, of coarse,
$\omega$-independent). 

For a generic initial data in the time domain, one is able to write
these data as a sum over $l, m$, and integration over $\omega$.
Namely, $\psi(r,\theta,\varphi,t)= \sum\limits_{m= -\infty}^{\infty}
\sum\limits_{l= -m}^{m} e^{im \varphi} \int_{-\infty}^{\infty}
R_l^m(r,t) S_l^m(\theta,a\omega) e^{-i \omega t}d\omega$ with $t=0$.
Then, each mode [characterized by a particular spheroidal wave
function $S_{l}^{m}(\theta,a\omega)$] evolves {\it independently}.
Nevertheless, Eq. (\ref{Eq20}) implies that each spheroidal wave
function can be expressed as a sum over the ($\omega$-independent)
spherical harmonics (were the $\omega$-dependence of the spheroidal
wave function is now moved to the expansion coefficients). That is,
the initial data can be expressed also as $\psi(r,\theta,\varphi,t)=
\sum\limits_{m= -\infty}^{\infty} \sum\limits_{l= -m}^{m}
R_l^{'m}(r,t) Y_l^m(\theta)e^{im \varphi}$ with $t=0$.

It is convenient to use the basis of the  ($\omega$-dependent) spheroidal wave
functions and carry out the analysis in the frequency domain; 
these functions allow us to separate the
$\theta$-dependence from the radial dependence, and, more important,
allow us to separate the different modes (different values of $l$) so 
each mode evolves {\it independently}, see Eq. (12) of
\cite{Hod}. However, since our goal is to calculate 
the {\it temporal} dependence of the field [in terms of the time $t$
(or $u, v$)], the final expression for the field 
should not involve the $\omega-${\it dependent} spheroidal
harmonics. Therefore, the 
final expression for the field is most naturally expressed in terms of 
the $\omega$-{\it independent} spherical harmonics.

We shall consider two kinds of initial data: A generic initial
data, which can be decomposite in terms of the basis of the 
spheroidal wave functions. Then each spheroidal wave function evolves
independently. In view of Eq. (\ref{Eq20}), 
this corresponds to the assumption that the (generic)
initial pulse consists of all allowed modes (all spherical harmonics
functions with $l \geq |m|$). 

On the other hand, in order to compare
our {\it analytical} results with the {\it numerical} results of
Krivan et. al. \cite{Krivan1}, one should consider an initial data
which is made of a single mode. This corresponds to the assumption 
that the initial angular distribution is characterized by a 
particular spherical harmonic function $Y_{l^*}^m$. 
We begin with this case, which
can be compared with the numerical results of \cite{Krivan1}.

\section{Pure initial Data}\label{Sec4}

\subsection{Asymptotic behaviour at timelike infinity}\label{Sec4A}

We consider first the 
behaviour of the scalar field at the asymptotic region of {\it timelike infinity} $i_+$.
As was explained, the late-time behaviour of the field should follow from the
{\it low}-frequency contribution to the Green's function. Actually, it is easy 
to verify that the effective contribution to the integral in 
Eq. (\ref{Eq25}) should come from $|\omega|$=$O({1 \over t})$. 
Thus, in order to obtain the asymptotic
behaviour of the field at {\it timelike infinity} (where $y,y' \ll t$),
we may use the $|\omega|r \ll 1$ limit of $\tilde \Psi_1(r,\omega)$. Using
Eq. 13.5.5 of \cite{Abram} one finds

\begin{equation}\label{Eq26}
\tilde \Psi_1(r,\omega) \simeq Ar^{l +1}\  .
\end{equation}
Substituting this in Eq. (\ref{Eq25}) we obtain

\begin{eqnarray}\label{Eq27}
G^C(z,z';t)& =&  \sum\limits_{l=|m|}^{\infty} 
{{2i M  } \over {\pi  [(2l+1)!!]^2}} (yy')^{l+1} \nonumber \\
&& \int_{0}^{-i \infty} {S_{l}(\theta,a\omega)} {S_{l}(\theta',a\omega)}
\omega ^{2l+2} e^{-i \omega t} d \omega\  .
\end{eqnarray}

Using the representation for the spheroidal wave functions
$S_{l}$, Eq. (\ref{Eq20}), togather with the fact that the integral Eq. (\ref{Eq23})
vanishes unless $l=k, k \pm 2$, we find that the asymptotic late-time
behaviour of the $l$ mode (where $|l^*-l|$ is even and $l \geq |m|$)
is dominated by the following effective Green's function:

\begin{eqnarray}\label{Eq28}
G^C_l(z,z';t)& =&  \sum\limits_{k=|m|+p}^{L} 
{{2i M  } \over {\pi  [(2k+1)!!]^2}} (yy')^{k+1} \nonumber \\
&& C_{kl}C_{kl^*-q} Y_{l}(\theta) Y_{l^*-q}^{*}(\theta')
\int_{0}^{-i \infty}  (a\omega)^{l^*+l-2k-q} \omega ^{2k+2} e^{-i \omega t} d \omega\  ,
\end{eqnarray}
where $p=0$ if $l^*-|m|$ is even and $p=1$ 
otherwise, and $q=2$ if $l^* \geq |m|+2$ and $q=0$ otherwise.
Here, $L=l^*-q$ for $l \geq l^*$ modes, and $L=l$ for $l \leq l^*-2$ modes.
Performing the integration, we obtain

\begin{eqnarray}\label{Eq29}
G^C_l(z,z';t)& =&  \sum\limits_{k=|m|+p}^{L} 
{{2M(-1)^{(l^*+l+2-q)/2} (l^*+l+2-q)!  } \over {\pi  [(2k+1)!!]^2}} (yy')^{k+1} \nonumber \\
&& C_{kl}C_{kl^*-q} Y_{l}(\theta) Y_{l^*-q}^{*}(\theta')
  a^{l^*+l-2k-q} t^{-(l^*+l+3-q)}\  .
\end{eqnarray}
Thus, the late-time behaviour of a scalar field at the
asymptotic region of timelike infinity $i_+$ is dominated by the
lowest allowed mode, i.e., by the $l=|m|$ mode if $l^*-|m|$ is
even, and by the $l=|m|+1$ mode if $l^*-|m|$ is odd. The corresponding
damping exponent is $-(l^*+|m|+p+1)$ if $l^* \geq |m|+2$, and
$-(2l^*+3)$ if $l^*=|m|, |m|+1$.

These {\it analytical} results should be compared with the {\it
  numerical} results of Krivan et. al. \cite{Krivan1}. The comparison
can be made for the specific initial data considered in
\cite{Krivan1}, and is presented in Table \ref{Tab3} -- we find a
perfect agreement between our analytical results and the numerical
results of Krivan et. al. \cite{Krivan1}.

\subsection{Asymptotic behaviour at future null infinity}\label{Sec4B}

We further consider the behaviour of the scalar field at the
asymptotic region of future null infinity $scri_+$.
It is easy to verify that for this case
the effective frequencies contributing to the integral in Eq. (\ref{Eq25}) are
of order $O({1 \over u})$.
Thus, for $y-y' \ll t \ll 2y-y'$ one may use the
$|\omega |y' \ll 1$ asymptotic limit for $\tilde \Psi_1(y',\omega)$
and the $|\omega |y \gg 1$
($Im \omega < 0$) asymptotic limit of $\tilde \Psi_1(y,\omega)$. Thus,

\begin{equation}\label{Eq30}
\tilde \Psi_1(y',\omega) \simeq Ay'^{l +1}\  ,
\end{equation}
and

\begin{equation}\label{Eq31}
\tilde \Psi_1(y,\omega) \simeq Ae^{i \omega y} (2l+1)! {{e^{-i {\pi \over 2}
(l+1-2i\omega M)} (2\omega)^{-l-1+2i\omega M} y^{2i\omega M}} \over
{\Gamma (l+1+2i\omega M)}}\  ,
\end{equation}
where we have used Eqs. 13.5.5 and 13.5.1 of \cite{Abram}, respectively.
Substituting this in Eq. (\ref{Eq25}) we obtain (using the $M\omega \to 0$
limit)

\begin{eqnarray}\label{Eq32}
G^C(z,z';t)& =&  \sum\limits_{l=|m|}^{\infty} 
{{M2^{-l} (2l+1)! (-1)^{(l+2)/2} } \over {\pi l! [(2l+1)!!]^2}} y'^{l+1} \nonumber \\
&& \int_{0}^{-i \infty} {S_{l}(\theta,a\omega)} {S_{l}(\theta',a\omega)}
\omega ^{l+1} e^{-i \omega(t-y)} d \omega\  .
\end{eqnarray}
Using the representation Eq. (\ref{Eq20}) for the spheroidal wave functions
$S_{l}$, togather with the fact that the integral Eq. (\ref{Eq23})
vanishes unless $l=k,k \pm 2$, we find that the 
behaviour of the $l$ mode (where $|l^*-l|$ is even and $l \geq |m|$)
at the asymptotic region of null infinity $scri_+$ 
is dominated by the following effective Green's function:

\begin{eqnarray}\label{Eq33}
G^C_l(z,z';t)& =&  \sum\limits_{k=l^*-q_1}^{l^*+q_2} 
{{M2^{-k} (2k+1)! (-1)^{(k+2)/2} } \over {\pi k!
    [(2k+1)!!]^2}} y'^{k+1} \nonumber \\
&& C_{kl} Y_{l}(\theta) Y_{k}^{*}(\theta')  \int_{0}^{-i \infty}
(a\omega)^{l-k}  \omega ^{k+1} e^{-i \omega(t-y)} d \omega\  ,
\end{eqnarray}
where $q_2=0$ for the $l=l^*$ mode and $q_2=2$ for $l \geq l^*+2$
modes ($q_1=2$ if $l^* \geq |m|+2$ and $q_1=0$ otherwise), and

\begin{eqnarray}\label{Eq34}
G^C_l(z,z';t)& =&  
{{M2^{-l} (2l+1)! (-1)^{(l+2)/2} } \over {\pi l! [(2l+1)!!]^2}} y'^{l+1} \nonumber \\
&& C_{ll^*-2} Y_{l}(\theta) Y_{l^*-2}^{*}(\theta')  \int_{0}^{-i \infty}
(a\omega)^{l^*-2-l} \omega ^{l+1} e^{-i \omega(t-y)} d \omega\  ,
\end{eqnarray}
for $l \leq l^*-2$ modes.
Performing the integration in Eqs. (\ref{Eq33}) and (\ref{Eq34}), one finds

\begin{eqnarray}\label{Eq35}
G^C_l(z,z';t)& =&  \sum\limits_{k=l^*-q_1}^{l^*+q_2} 
{{M2^{-k} (-1)^{(l+k+2)/2}  (2k+1)! (l+1)!} \over {\pi k! [(2k+1)!!]^2}} y'^{k+1} \nonumber \\
&& C_{kl} Y_{l}(\theta) Y_{k}^{*}(\theta') 
a^{l-k}  u^{-(l+2)}\  ,
\end{eqnarray}
for $l \geq l^*$ modes, and

\begin{eqnarray}\label{Eq36}
G^C_l(z,z';t)& =&  
{{M2^{-l}(-1)^{(l^*+l)/2} (l^*-1)!(2l+1)! } \over {\pi l! [(2l+1)!!]^2}} y'^{l+1} \nonumber \\
&& C_{ll^*-2} Y_{l}(\theta) Y_{l^*-2}^{*}(\theta') 
a^{l^*-2-l}  u^{-l^*} \  ,
\end{eqnarray}
for $l \leq l^*-2$ modes.
Thus, we find that the late-time behaviour of a scalar field at the
asymptotic region of null infinity $scri_+$ is dominated by the
$l \leq l^*-2$ modes provided that $l^* \geq |m|+2$ 
(where $|l^*-l|$ is even 
and $l \geq |m|$). These modes decay as an inverse power of
the retarded time $u$, with a damping exponent equals to $-l^*$. 
If $l^*=|m|, |m|+1$ then the late-time
evolution will be dominated by the $l=l^*$ mode, which has a damping
exponent equals to $-(l^*+2)$.

\subsection{Asymptotic behaviour at the black-hole outer horizon}
\label{Sec4C}

Finally, we consider the behaviour of the field at the
black-hole outer-horizon $H_+$. The asymptotic solution at the horizon
($y \to -\infty$) is given by  \cite{Teukolsky2} (see Ref. \cite{Hod} 
for additional details)

\begin{equation}\label{Eq37}
\tilde \Psi_1(y,\omega)=C(\omega) e^{-i(w-mw_+)y}\  ,
\end{equation}
where $w_+=a/(2Mr_+)$. In addition, 
we use Eq. (\ref{Eq30}) for $\tilde \Psi_1(y',\omega)$. 

Using the representation Eq. (\ref{Eq20}) for the spheroidal wave functions
$S_{l}$, togather with the fact that the integral Eq. (\ref{Eq23})
vanishes unless $l=k, k \pm 2$, we find that the asymptotic 
behaviour [where $C(\omega)=const. +O(M\omega)$] 
of the $l$ mode (where $|l^*-l|$ is even and $l \geq |m|$)
at the black-hole outer horizon $H_+$ 
is dominated by the following effective Green's function:

\begin{eqnarray}\label{Eq38}
G^C_l(z,z';t)& =&  \sum\limits_{k=|m|+p}^{L} 
\Gamma_k {{2M(-1)^{(l^*+l+2-q)/2} (l^*+l+2-q)!  } 
\over {\pi  [(2k+1)!!]^2}} y'^{k+1} \nonumber \\
&& C_{kl}C_{kl^*-q} Y_{l}(\theta) Y_{l^*-q}^{*}(\theta')
  a^{l^*+l-2k-q} e^{imw_+y} v^{-(l^*+l+3-q)}\  ,
\end{eqnarray}
where $q, p$ and $L$ are defined as before, and $\Gamma_k$ are constants.
Hence, we find that the late-time behaviour of a scalar field at
the black-hole outer horizon is dominated by the
lowest allowed mode, i.e., by the $l=|m|$ mode if $l^*-|m|$ is
even, and by the $l=|m|+1$ mode if $l^*-|m|$ is odd.

\section{Generic Initial Data}\label{Sec5}

In this section we consider the generic case. That is, we assume that the initial
pulse consists of all the allowed ($l \geq |m|$) modes (see also the
recent analysis of Barack and Ori \cite{BarOrl}). 
The analysis here is very similar to the one presented in \ref{Sec4}. 
We consider first the 
behaviour of the scalar field at the asymptotic 
region of {\it timelike infinity} $i_+$. Using Eq. (\ref{Eq27}),
togather with the representation Eq. (\ref{Eq20}) for the spheroidal
wave functions, we find that the asymptotic late-time
behaviour of the $l$ mode (where $l \geq |m|$)
is dominated by the following effective Green's function:

\begin{eqnarray}\label{Eq39}
G^C_l(z,z';t)& =&  
{{2i M  } \over {\pi  [(2k+1)!!]^2}} (yy')^{k+1} \nonumber \\
&& C_{kl} Y_{l}(\theta) Y_{k}^{*}(\theta')
\int_{0}^{-i \infty}  (a\omega)^{l-k} \omega ^{2k+2} e^{-i \omega t} d \omega\  ,
\end{eqnarray}
where $k=|m|+p$, and $p=0$ if $l-|m|$ is even, and $p=1$ otherwise. 
Performing the integration, we obtain

\begin{eqnarray}\label{Eq40}
G^C_l(z,z';t)& =&  
{{2M(-1)^{(l+|m|+p+2)/2} (l+|m|+p+2)!  } \over {\pi  [(2|m|+2p+1)!!]^2}} (yy')^{|m|+p+1} 
\nonumber \\
&& C_{|m|+pl} Y_{l}(\theta) Y_{|m|+p}^{*}(\theta')
  a^{l-|m|-p} t^{-(l+|m|+p+3)}\  .
\end{eqnarray}
We emphasize that the power indices $l+|m|+p+3$ (in absolute value)
in {\it rotating} Kerr spacetimes are {\it smaller} than the
corresponding power indices (the well known $2l+3$)
in {\it spherically} symmetric Schwarzschild 
spacetimes. (There is an equality only for the $l=|m|,|m|+1$ modes). 
This implies a {\it slower} decay of perturbations
in rotating Kerr spacetimes. From Eq. (\ref{Eq40}) it is easy to see
that the time scale $t_c$ at which the late-time tail of rotating
gravitational collapse is considerably different from the corresponding tail of non
rotating collapse (for the $l > |m|+1$ modes) is $t_c=yy'/a$, where
$y'$ is roughly the average location of the initial pulse.

We further consider the asymptotic behaviour of the field 
at future null infinity $scri_+$. Using Eq. (\ref{Eq32}), 
togather with the representation Eq. (\ref{Eq20}) for the spheroidal
wave functions, one finds that the behaviour of the 
$l$ mode at the asymptotic region of null infinity is dominated 
by the following effective Green's function:

\begin{eqnarray}\label{Eq41}
G^C_l(z,z';t)& =&  \sum\limits_{k=|m|+p}^{l} 
{{M2^{-k} (2k+1)! (-1)^{(k+2)/2} } \over {\pi k!
    [(2k+1)!!]^2}} y'^{k+1} \nonumber \\
&& C_{kl} Y_{l}(\theta) Y_{k}^{*}(\theta')  \int_{0}^{-i \infty}
(a\omega)^{l-k}  \omega ^{k+1} e^{-i \omega(t-y)} d \omega\  ,
\end{eqnarray}
where $p$ is defined above. 
Performing the integration we obtain

\begin{eqnarray}\label{Eq42}
G^C_l(z,z';t)& =&  \sum\limits_{k=|m|+p}^{l} 
{{M2^{-k} (-1)^{(l+k+2)/2}  (2k+1)! (l+1)!} \over {\pi k! [(2k+1)!!]^2}} y'^{k+1} \nonumber \\
&& C_{kl} Y_{l}(\theta) Y_{k}^{*}(\theta') 
a^{l-k}  u^{-(l+2)}\  .
\end{eqnarray}

Finally, we consider the behaviour of the field at the
black-hole outer-horizon $H_+$. Following an analysis similar
to the one presented in Sec. \ref{Sec4C}, 
we find that the asymptotic 
behaviour of the $l$ mode (where $l \geq |m|$) at 
the black-hole outer horizon $H_+$ 
is dominated by the following effective Green's function:

\begin{eqnarray}\label{Eq43}
G^C_l(z,z';t)& =&  
\Gamma_k {{2M(-1)^{(l+|m|+p+2)/2} (l+|m|+p+2)!  } \over {\pi  [(2|m|+2p+1)!!]^2}} y'^{|m|+p+1} 
\nonumber \\
&& C_{|m|+pl} Y_{l}(\theta) Y_{|m|+p}^{*}(\theta')
  a^{l-|m|-p}e^{imw_+y} v^{-(l+|m|+p+3)}\  .
\end{eqnarray}

\section{Summary of Results}\label{Sec6}

The asymptotic late-time evolution of a test scalar field on a
realistic Kerr background is
characterized by inverse power-law tails at the three asymptotic
regions: timelike infinity $i_{+}$, future null infinity $scri _{+}$, 
and the black-hole outer-horizon $H_{+}$ (where the power-law
behaviour is multiplied by an oscillatory term, caused by the dragging
of reference frames at the event horizon). This late-time behaviour 
is similar to the the late-time evolution of massless fields on a 
{\it spherically}-symmetric Schwarzschild background, originally
analyzed by Price \cite{Price}.

However, there is one important feature (besides the oscillatory
behaviour along the black-hole horizon) which is unique to {\it rotating}
collapse: Active {\it coupling} of different modes. Physically,
this phenomena is caused by the dragging of reference frames, due to
the black-hole (or star's) rotation (this phenomena is absent in the
non-rotating $a=0$ case). Hence, the late-time tails of realistic
rotating collapse are dominated by a mode, which, in
general, is different from the initial one.

As shown in our analysis, there are various cases, which
result in different types of asymptotic late-time evolutions
(different damping exponents). These
distinct cases are characterized by the values of $l^*$ 
and $|m|$, and the parity of $l^*-|m|$. 
The dominant modes at asymptotic late-times 
and the values of the corresponding damping exponents (for pure
initial data, characterized by a particular spherical harmonic
function $Y_{l^*}^m$) are summarized in Tables \ref{Tab1} (for $l^* \geq |m|+2$) 
and \ref{Tab2} (for $l^*=|m|,|m|+1$). These {\it analytical} results
should be compared with the {\it numerical} results
of \cite{Krivan1} -- 
we find a complete {\it agreement} between the two, as can be seen 
in Table \ref{Tab3}.

In summary, the rotation of a black hole (or a star) results in an 
active coupling of different modes. Thus, new
modes, which are different from the original one, 
would be generated during a rotating collapse and
may play an important role at asymptotic late times. Hence, the radiative tail of
{\it rotating} gravitational collapse decays {\it slower} than the
corresponding tail of a non-rotating collapse. In general, 
a rotating Kerr black hole becomes
`bald' {\it slower} than a spherically-symmetric Schwarzschild
black hole. 

\bigskip
\noindent
{\bf ACKNOWLEDGMENTS}
\bigskip

I thank Tsvi Piran for discussions. 
This research was supported by a grant from the Israel Science Foundation.

\appendix

\section{Coupling of the real and imaginary parts of $\Psi$}\label{SecApp}

In general, the $\Psi^m$ functions are complex quantities. Owing to the
purely real character of the $B_1(r,\theta)$ coefficient, the real
and imaginary parts of each $\Psi$ are decoupled for 
axially-symmetric ($m=0$) modes (where $B_2$ vanishes and the Green's
function is purely real).
However, for non-axially symmetric ($m \neq 0$) modes, the {\it imaginary} coefficient
$B_2(r)$ {\it mixes} the real and imaginary parts of the {\it same} mode.
Hence, the real part of a particular mode will always generate the
corresponding imaginary part of the same mode, and vice versa
(note, however, that the coefficient $B_2(r)$ is
$\theta-$independent. Hence, contrasted with the $B_1(r,\theta)$ coefficient, 
it cannot mix nearest neighbors modes). 
We now assume that the initial data for $\Psi$ is purely real,
and (without loss of generality) we assume that its angular distribution 
is characterized as before by a particular spherical harmonic 
function $Y_{l^*}^m$.

We consider first the asymptotic
behaviour of the imaginary part of $\Psi^m$ at {\it timelike infinity} $i_+$.
Following an analysis similar to the one presented in Sec. \ref{Sec4A},
one finds that the asymptotic late-time behaviour of the imaginary
part of the $l$ mode (where $|l^*-l|$ is even and $l \geq |m| > 0$) is dominated
by the following effective Green's function:

\begin{eqnarray}\label{EqA1}
G^C_l(z,z';t)& =&  \sum\limits_{k=|m|+p}^{L} 
{{2M(-1)^{(l^*+l+2)/2} (l^*+l+2)!  } \over {\pi  [(2k+1)!!]^2}} (yy')^{k+1} \nonumber \\
&& C_{kl}C_{kl^*} Y_{l}(\theta) Y_{l^*}^{*}(\theta')
  a^{l^*+l-2k} t^{-(l^*+l+3)}\  ,
\end{eqnarray}
where $p=0$ if $l^*-|m|$ is even and $p=1$ 
otherwise. Here, $L=l^*$ for $l \geq l^*$ modes and $L=l$ 
for $l \leq l^*-2$ modes. Hence, we find that the evolution at 
timelike infinity $i_+$ of the imaginary part of $\Psi$ is dominated by the
lowest allowed mode.

We further consider the asymptotic behaviour of the imaginary part of
$\Psi$ at future null infinity $scri_+$. Following an analysis similar
to the one presented in Sec. \ref{Sec4B},
one finds that the asymptotic late-time behaviour of the imaginary
part of the $l$ mode is dominated
by the following effective Green's function:

\begin{eqnarray}\label{EqA2}
G^C_l(z,z';t)& =&  
{{M(-1)^{(l+l^*+2)/2} 2^{-l^*} (2l^*+1)! (l+1)! } \over {\pi l^*!
    [(2l^*+1)!!]^2}} y'^{l^*+1} \nonumber \\
&& C_{l^*l} Y_{l}(\theta) Y_{l^*}^{*}(\theta') 
a^{l-l^*}  u^{-(l+2)}\  ,
\end{eqnarray}
for $l \geq l^*$ modes, and

\begin{eqnarray}\label{EqA3}
G^C_l(z,z';t)& =&  
{{M(-1)^{(l^*+l+2)/2} 2^{-l} (2l+1)!(l^*+1)! } \over {\pi l! [(2l+1)!!]^2}} y'^{l+1} \nonumber \\
&& C_{ll^*} Y_{l}(\theta) Y_{l^*}^{*}(\theta') 
a^{l^*-l}  u^{-(l^*+2)} \  ,
\end{eqnarray}
for $l \leq l^*-2$ modes.
Thus, we find that the evolution at null infinity $scri_+$ 
of the imaginary part of $\Psi$ is dominated by the
$l \leq l^*$ modes (provided that $|l-l^*|$ is even 
and $l \geq |m| > 0$).

Finally, we note that the Green's function Eq. (\ref{Eq38}) is already 
a complex function for non-axially symmetric ($m \neq 0$) modes. 
Thus, the real and imaginary parts of the {\it same} mode are
coupled by both the $B_1(r,\theta)$ and $B_2(r)$ coefficients. Since 
the coupling can occur by the $\theta-$dependent
coefficient $B_1(r,\theta)$ (which couples nearest neighbors
modes), the asymptotic 
behaviour at the black-hole outer horizon $H_+$
of the imaginary part of the $l$ mode 
is dominated by the effective Green's function Eq. (\ref{Eq38}).

\begin{table}
\caption{Dominant modes and asymptotic damping exponents for $l^*
  \geq |m|+2$}
\label{Tab1}
\begin{tabular}{lcc}
asymptotic region & dominant mode(s)& damping exponent\\
\tableline
timelike infinity & $|m|+p$ & $-(l^*+|m|+p+1)$ \\
null infinity & $|m|+p \leq l \leq l^*-2$ & $-l^*$ \\
outer horizon & $|m|+p$ & $-(l^*+|m|+p+1)$ \\
\end{tabular}
\end{table}

\begin{table}
\caption{Dominant mode and asymptotic damping exponents 
for $l^*=|m|,|m|+1$}
\label{Tab2}
\begin{tabular}{lcc}
asymptotic region & dominant mode& damping exponent\\
\tableline
timelike infinity & $l^*$ & $-(2l^*+3)$ \\
null infinity & $l^*$ & $-(l^*+2)$ \\
outer horizon & $l^*$ & $-(2l^*+3)$ \\
\end{tabular}
\end{table}

\begin{table}
\caption{analytical vs. numerical results -- the dominant damping exponent at $i_+$}
\label{Tab3}
\begin{tabular}{lcc}
initial data & analytical result& numerical result\\
\tableline
$l=m=1$& $-5$ & $-4.87$ \\
$l=m=2$& $-7$ & $-6.98$ \\
$l=2, m=0$& $ -3$ & $ \sim -2.9$ \\
\end{tabular}
\end{table}

\end{document}